\documentclass[12pt]{article}
\usepackage{graphicx}
\usepackage{amsmath}
\usepackage{hyperref}
\usepackage{multirow}
\usepackage{booktabs}
\usepackage{INTERSPEECH2021}

\title{Text-Dependent Speaker Verification (TdSV) Challenge 2024: Team Naïve System Report }

\name{Amir Mohammad Rostami$^1$, Pourya Jafarzadeh$^1$}

\address{
  $^1$Self-Organized and Independent Participants}
\email{amirmohammadrostami@yahoo.com   pouryajafarzadeh@gmail.com}

\date{}

\begin{document}

\maketitle

\begin{abstract}
This paper presents a system for the 2024 Text-Dependent Speaker Verification (TdSV) Challenge. The system achieved a Minimum Detection Cost Function (MinDCF) of 0.0461 and an Equal Error Rate (EER) of 1.3\%.
Our approach focused on adapting existing state-of-the-art neural networks, ResNet-TDNN and NeXt-TDNN, originally trained on the VoxCeleb dataset. This strategy was chosen because of the limited challenge duration and the available resources at the time. In addition, we designed a lightweight and resource-efficient model, EfficientNet-A0, trained specifically on the challenge dataset to improve adaptation and strengthen the ensemble approach.
Our system combines advanced neural architectures, extensive data augmentation, and optimised hyperparameters. These components helped achieve strong performance in text-dependent speaker verification. The results also demonstrate the effectiveness of multi-model ensemble learning for both speaker and phrase verification.
\end{abstract}

\section{Introduction}

\begin{figure*}[ht]
    \centering
    \includegraphics[width=\linewidth]{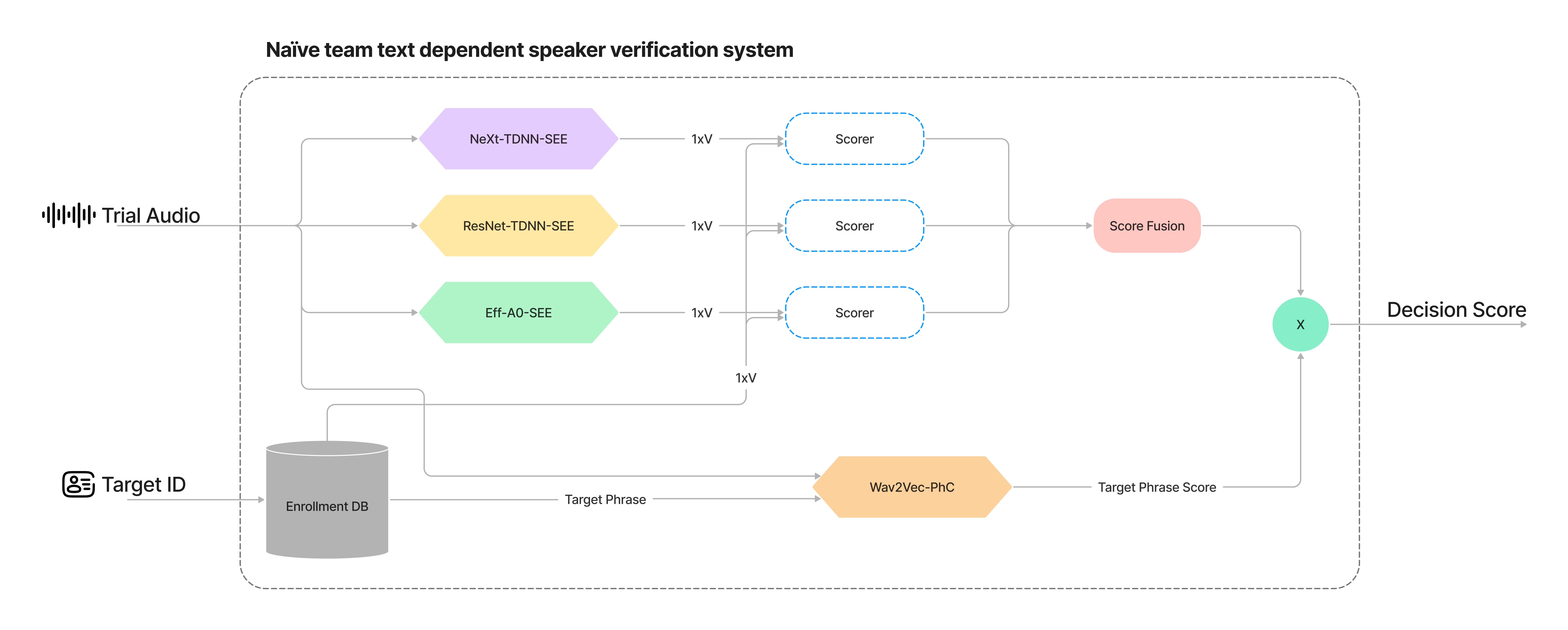}
    \caption{Naïve team text-dependent speaker verification system architecture. The system consists of three Speaker Embedding Extractors (SEEs), a wav2vec 2.0 Phrase Classifier (PhC), individual scorers, and a score fusion mechanism.}
    \label{fig:system_architecture}
\end{figure*}

Text-Dependent Speaker Verification (TdSV) plays a crucial role in biometric security systems, particularly in applications where both speaker identity and the spoken content are essential for authentication. Unlike Text-Independent Speaker Verification (TiSV), where the verification task can be performed with arbitrary speech content, TdSV mandates the speaker to pronounce specific or prompted phrases. This introduces an additional challenge, as the system must accurately recognise the speaker’s identity while also verifying the correctness of the spoken phrase, thereby integrating speaker recognition and speech content verification in tandem \cite{heigold2016end, koshinaka2019advances}.\\
The primary advantage of TdSV systems is their ability to enhance security through the fusion of both speaker characteristics and linguistic content. This dual requirement makes TdSV systems more robust against spoofing attacks and phrase-based replay attacks, a key advantage over TiSV methods \cite{alcheikh2021survey}. However, this task is inherently more complex due to the need for precise modelling of both voice and phrase-specific patterns, leading to a more demanding requirement for feature extraction and model design \cite{chen2021text}.

The 2024 Text-Dependent Speaker Verification (TdSV) Challenge\cite{tdsvc2024plan} provided an opportunity to push the boundaries of current state-of-the-art systems. The challenge encouraged participants to explore innovative methods for handling the intricacies of both speaker identification and phrase detection, leveraging recent advances in deep learning, speaker embedding techniques, and data augmentation strategies. In recent years, architectures such as Time Delay Neural Networks (TDNNs) \cite{snyder2018xvector}, convolutional neural networks (CNNs) \cite{villalba2019state}, and transformer-based models \cite{yu2021dual} have become pivotal in achieving high accuracy in speaker verification tasks. The integration of these techniques has further improved the ability of TdSV systems to generalise across different recording conditions and linguistic variations \cite{zhang2020deep}.

\begin{figure}[h!]
    \centering
    \includegraphics[width=0.95\columnwidth]{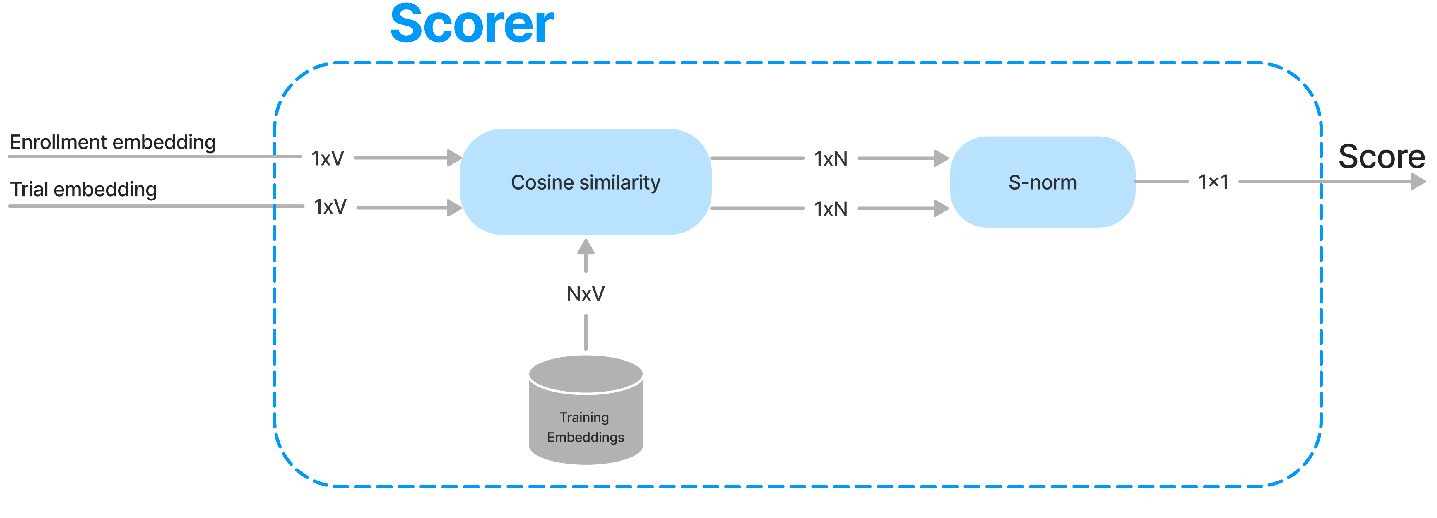}
    \caption{Scorer module architecture, showing the cosine similarity calculation between enrollment and trial embeddings, with S-norm normalisation based on the training embedding database.}
    \label{fig:scorer_module}
\end{figure}

In this paper, we present the text-dependent speaker verification system of the \textit{Naïve} team,  which combines several state-of-the-art neural network architectures for robust speaker verification. Due to the challenge constraints, including a limited timeline of nine weeks and no dedicated GPU access, we focused on adapting pretrained models.
Our approach used two pretrained neural networks, ResNet-TDNN and NeXt-TDNN, trained on the VoxCeleb dataset to make the best use of the available resources and limited time. In parallel, we trained a lightweight and resource-efficient model, EfficientNet-A0, on the challenge dataset to improve adaptation to the challenge requirements and strengthen the ensemble performance in low-resource settings.
This combination of fine-tuning pretrained models and introducing a lightweight model allowed the ensemble approach to perform well across different verification tasks. As shown in Figure~\ref{fig:system_architecture}, the system is designed to verify speakers by analysing both the speaker’s voice characteristics and the spoken content. The verification process consists of two main tasks: speaker verification and phrase verification.\\
The system integrates three different deep neural networks, NeXt-TDNN-SEE~\cite{next}, ResNet-TDNN-SEE~\cite{resnet}, and EfficientNet-A0-SEE~\cite{eff}, where each model acts as a Speaker Embedding Extractor (SEE). These models process the trial audio and generate high-dimensional speaker embeddings (1xV) that capture the unique characteristics of a speaker’s voice.
The generated embeddings are then passed to a dedicated Scorer module, which compares the trial embeddings with the enrolled speaker identity stored in a database of enrollment embeddings.\\
The Scorer module, illustrated in Figure~\ref{fig:scorer_module}, plays a pivotal role in the verification process. It computes the similarity between the enrollment and trial embeddings using cosine similarity, a widely adopted metric in speaker verification that measures the angle between two vectors in the embedding space, providing a normalised measure of proximity between the embeddings. To further refine the scoring process, the system applies S-norm normalisation, which adjusts the raw similarity score based on the distribution of speaker embeddings in the training database. The normalised score is computed as follows:

\begin{multline}
\text{Score}_{\text{S-norm}}(e, t) = \frac{\text{Score}(e, t) - \mu_{\text{trial}}}{\sigma_{\text{trial}}} \\
+ \frac{\text{Score}(e, t) - \mu_{\text{enroll}}}{\sigma_{\text{enroll}}}
\end{multline}

where:
\begin{itemize}
    \item \(\text{Score}(e, t)\) is the raw cosine similarity between the enrollment embedding \(e\) and the trial embedding \(t\).
    \item \(\mu_{\text{trial}}\) and \(\sigma_{\text{trial}}\) represent the mean and standard deviation of the trial scores obtained against a cohort of enrollment embeddings from the training dataset.
    \item \(\mu_{\text{enroll}}\) and \(\sigma_{\text{enroll}}\) represent the mean and standard deviation of the enrollment scores obtained against a cohort of trial embeddings from the training dataset.
\end{itemize}
This scoring mechanism ensures that the final similarity score is robust, compensating for potential variations in speaker representations caused by different environmental conditions, such as noise or varying recording settings.

In addition to speaker verification, the system incorporates a Phrase Verification module, which verifies that the spoken phrase matches the enrolled target phrase. This is achieved through the Wav2vec-PhC (Phrase Classifier) module. The wav2vec 2.0 \cite{baevski2020wav2vec} model serves as a powerful feature extractor for the spoken content, and the extracted features are fed into fully connected layers for classification of the phrase into predefined categories. The phrase score is then combined with the speaker score within a multi-task verification framework.

The scores from the speaker verification module and the phrase verification module are fused to generate the final decision score. This score fusion ensures that the system makes decisions based on both who is speaking and what is being said, significantly enhancing its reliability in text-dependent speaker verification tasks.

The system's architecture capitalises on the complementary strengths of each model in the ensemble and the robust scoring mechanism to improve overall performance. By leveraging the ensemble learning approach, the system benefits from the individual strengths of the NeXt-TDNN, ResNet-TDNN, and EfficientNet models, yielding improved generalisation across various speaker verification scenarios.

The proposed system offers a comprehensive solution that integrates multi-model speaker verification with content verification, making it particularly well-suited for text-dependent speaker verification tasks. The use of advanced deep learning architectures, a robust scoring mechanism, and scoring fusion techniques led to strong performance in this challenging domain.


\section{Dataset}

The evaluation dataset for this challenge was derived from the DeepMine corpus\cite{deepmine2018odyssey,deepmine2019asru}, a versatile and multi-purpose dataset that includes a text-dependent speaker verification subset (Part 1). The corpus contains utterances based on 10 predefined phrases, recorded under controlled conditions to minimise external variables. Each speaker is required to repeat a given passphrase three times, ensuring a balanced representation of speaker and content variation for the speaker verification task.

\subsection{Training Condition}

The training condition for this challenge was defined as the amount of data and resources used to build the Speaker Recognition (SR) system. We adopted a fixed training condition, where participants were required to train their systems solely using a designated set of datasets. The fixed training set consists of the following:

\begin{itemize}
    \footnotesize 
    \item VoxCeleb 1 \& 2
    \item LibriSpeech
    \item Mozilla Common Voice Farsi
    \item DeepMine
\end{itemize}

The data was recorded in real-life environments in Iran, incorporating various background noise types to simulate real-world conditions. Persian (Farsi) serves as the primary language of the dataset, although many participants also contributed to the English portion. Part 1 of the dataset consists of five Persian phrases and five English phrases, which were used in both text-dependent speaker verification tasks of the challenge. Table~\ref{tbl.phrases} lists the English phrases and their transliterations in Persian.

\begin{table}[tb]
    \renewcommand{\arraystretch}{1.2}
    \caption{Phrases in Task 1 of the challenge.}
    \centerline{
        \setlength\tabcolsep{12pt}
        \begin{tabular}{l | l }
            \toprule
            \midrule
            Id & Phrase \\
            \midrule
            01 & sedaye man neshandahandeye hoviyyate man ast. \\
            02 & sedaye har kas monhaser be fard ast. \\
            03 & hoviyyate man ra ba sedaye man tayid kon. \\
            04 & sedaye man ramze obure man ast. \\
            05 & baniadam azaye yekdigarand. \\
            06 & My voice is my password. \\
            07 & OK Google. \\ 
            08 & Artificial intelligence is for real. \\
            09 & Actions speak louder than words. \\
            10 & There is no such thing as a free lunch. \\
            \midrule
            \bottomrule
        \end{tabular}
    }
    \label{tbl.phrases}
\end{table}

\section{Methodology}

In this section, we describe the methodology used to achieve the results in the 2024 Text-Dependent Speaker Verification (TdSV) Challenge. The proposed approach integrates three different deep learning models, where each model contributes to the overall performance through advanced speaker embedding and training strategies.

\subsection{Model Architecture}

The proposed system utilises a dual-module architecture to effectively handle text-dependent speaker verification and phrase detection.

\subsubsection{Speaker Verification Module}
The speaker verification module employs an ensemble of three deep neural network models: NeXt-TDNN, ResNet TDNN, and EfficientNet-Absolute Zero. This ensemble approach allows for score fusion, enhancing overall performance by leveraging the strengths of each model.

\subsubsection{NeXt-TDNN}
The first model used is the NeXt-TDNN, an advanced architecture designed specifically for speaker verification tasks. This model was fine-tuned on the challenge dataset, leveraging its pre-trained weights obtained from the Vox1 and Vox2 datasets. 

The architecture of the NeXt-TDNN is summarised in Table \ref{tab:next_tdnn_architecture}. The NeXt-TDNN model employs a modernised multi-scale temporal convolution backbone, enhancing performance through a novel block design known as TS-ConvNeXt. This design replaces the traditional squeeze-and-excitation (SE) Res2Net block found in ECAPA-TDNN with a two-step multi-scale convolution structure. The TS-ConvNeXt block consists of two sub-modules: a temporal multi-scale convolution (MSC) module and a frame-wise feed-forward network (FFN). This architecture enables flexible capturing of inter-frame and intra-frame contexts, facilitating improved feature representation. 

We also implemented global response normalisation (GRN) in the FFN modules, which promotes selective feature propagation. Empirical evaluations indicate that the NeXt-TDNN model yields significant performance enhancements in speaker verification tasks while maintaining a compact parameter size and reduced inference time. 

The augmentation strategies, including spectral augmentation and mel-frequency feature extraction, significantly improve the model's robustness to various acoustic conditions, ensuring effective performance across diverse environments.

\begin{table}[h]
    \centering
    \caption{Architecture and Training Configuration of NeXt-TDNN}
    \label{tab:next_tdnn_architecture}
    \resizebox{\columnwidth}{!}{ 
        \begin{tabular}{|l|l|}
            \hline
            \textbf{Component} & \textbf{Value} \\ \hline
            \multirow{3}{*}{Convolutional Layers} & Depths: [3, 3, 3] \\ 
            & Dims: [256, 256, 256] \\ 
            & Kernel Sizes: [7, 65] \\ \hline
            Batch Size & 200 \\ \hline
            Learning Rate & $1 \times 10^{-6} \times \text{Batch Size} = 0.0002$ \\ \hline
            Embedding Size & 192 \\ \hline
            Max Frames per Sample & 400 \\ \hline
            Sampling Rate & 16000 Hz \\ \hline
            \multirow{2}{*}{Feature Extractor} & Type: Mel Transform \\ 
            & n\_fft: 512, win\_length: 400, \\ 
            & hop\_length: 160, n\_mels: 80, coef: 0.97 \\ \hline
            Spec Augmentation & Freq Mask Param: 8, Time Mask Param: 10 \\ \hline
            Optimizer & AdamW \\ \hline
            \multirow{2}{*}{Loss Function} & Type: Additive Angular Margin Softmax \\ 
            & Embedding Size: 192, Num Classes: 1958, Margin: 0.3, \\ 
            & Scale: 40 \\ \hline
            Max Epochs & 200 \\ \hline
            Scheduler & Cyclical Learning Rate \\ \hline
        \end{tabular}
    }
\end{table}

\subsubsection{ResNet TDNN}
In addition to the NeXt-TDNN, the ResNet TDNN model was fine-tuned on the challenge dataset. This model leverages residual connections, allowing it to effectively capture speaker characteristics through its deep learning architecture. The fine-tuning process was meticulously crafted, focusing on optimising hyperparameters to adapt effectively to the specific nuances of the challenge dataset.
The architecture and training configurations for the ResNet TDNN model are detailed in Table \ref{tab:resnet_tdnn_architecture}.\\
\begin{table}[h]
    \centering
    \caption{Architecture and Training Configuration of ResNet TDNN}
    \label{tab:resnet_tdnn_architecture}
    \resizebox{\columnwidth}{!}{%
    \begin{tabular}{|l|l|}
        \hline
        \textbf{Component} & \textbf{Value} \\ \hline
        Number of Epochs & 10 \\ \hline
        Batch Size & 10 \\ \hline
        Learning Rate & 0.0001 \\ \hline
        Base Learning Rate & $1 \times 10^{-8}$ \\ \hline
        Sentence Length & 3.0 seconds \\ \hline
        \multirow{3}{*}{Feature Parameters} & n\_mels: 80 \\ 
        & Left Frames: 0 \\ 
        & Right Frames: 0 \\ \hline
        \multirow{4}{*}{Embedding Model} & Input Size: 80 \\ 
        & Channels: [128, 128, 256, 256] \\ 
        & Strides: [1, 2, 2, 2] \\ 
        & Block Sizes: [3, 4, 6, 3] \\ \hline
        Embedding dimension & 256 \\ \hline
    \end{tabular}
    }
\end{table}The ResNet TDNN model benefits from its capacity to learn deep representations, thanks to its unique architecture, which incorporates several residual blocks. Each block enhances the model's ability to extract features effectively, facilitating a more robust understanding of speaker identities. The integration of batch normalisation and activation functions within the residual structure allows for improved convergence and model performance during training.\\
To further enhance the model's performance, we employed various augmentation strategies that help the model generalise better to different acoustic environments. These strategies include the addition of noise, reverberation, and random drops of frequency bands and temporal chunks. These augmentations introduce variability in the training data, allowing the model to become more resilient against real-world conditions that may vary significantly from the training environment.\\
The feature extraction process utilises a Mel-frequency bank approach, extracting 80 Mel-frequency cepstral coefficients (MFCCs) from each audio sample, which are crucial for capturing the spec

\subsubsection{EfficientNet-Absolute Zero}
To further enhance the system's robustness, the EfficientNet-Absolute Zero model trained from scratch using the challenge dataset. The EfficientNet-A0 architecture, developed based on the compound scaling method applied to EfficientNet-B0, provides a lightweight and efficient solution for deep learning tasks. Notably, this model was designed specifically for keyword spotting, emphasising performance efficiency and minimal computational overhead.\\
The architecture of the EfficientNet-Absolute Zero model includes critical parameters optimised for speaker verification tasks, making it a novel approach in this domain. The specifications for the embedding model and classifier are summarised in Table~\ref{tab:efficientnet_specifications}.\\
\begin{table}[h]
    \centering
    \caption{Specifications of EfficientNet-Absolute Zero}
    \label{tab:efficientnet_specifications}
    \begin{tabular}{|c|c|}
        \hline
        \textbf{Parameter} & \textbf{Value} \\
        \hline
        {Embedding Model Input Shape} & [301, 80] \\
        Width Coefficient & 0.35 \\
        Depth Coefficient & 0.25 \\
        Dropout Rate & 0.15 \\
        Drop Connect Rate & 0.25 \\
        Classifier Input Size & 256 \\
        Classifier Output Neurons & 1958 \\
        \hline
    \end{tabular}
\end{table}By training this model on the challenge dataset, we aimed to leverage its capabilities for recognizing predefined phrases with high accuracy, as well as generating speaker embeddings. This approach marks one of the first instances of applying EfficientNet-Absolute Zero to speaker verification tasks, to the best of our knowledge. The lightweight nature of this architecture, combined with its optimised design for the specific requirements of speaker verification, allows for efficient processing while maintaining high accuracy.\\
The EfficientNet-Absolute Zero model architecture not only improves performance metrics but also significantly reduces computational overhead, making it suitable for deployment in resource-constrained environments. By incorporating this model into our speaker verification system, we improve the robustness and reliability of the overall approach. This also provides a strong foundation for future work in this field.

\subsubsection{Phrase Detection Module}
In addition to the speaker verification module, a phrase detection module is incorporated to identify predefined phrases from the audio input. This module utilises the wav2vec 2.0 model, specifically the 
\texttt{lighteternal/wav2vec2-large-xlsr-53\\- greek} variant, which has been fine-tuned for speech classification tasks.\\
The architecture of the wav2vec model is summarised in Table \ref{tab:phrase_detection_architecture}.
\begin{table}[h]
    \centering
    \caption{Architecture Details of the Phrase Detection Module}
    \label{tab:phrase_detection_architecture}
    \resizebox{\columnwidth}{!}{%
    \begin{tabular}{|l|l|}
        \hline
        \textbf{Component} & \textbf{Value} \\ \hline
        Model Name & \texttt{lighteternal/wav2vec2-large-xlsr-53-greek} \\ \hline
        \multirow{3}{*}{Convolutional Layers} & Number of Layers: 7 \\ 
        & Kernel Sizes: [10, 3, 3, 3, 3, 2, 2] \\ 
        & Strides: [5, 2, 2, 2, 2, 2, 2] \\ \hline
        Attention Heads & 16 \\ \hline
        Hidden Layers & 24 \\ \hline
        Dropout Rates & Hidden Dropout: 0.1 \\ 
        & Attention Dropout: 0.1 \\ \hline
        Output Dimension (x-vector) & 512 \\ \hline
        \multirow{2}{*}{Fully Connected Layers} & Hidden Size: 1024 \\ 
        & Intermediate Size: 4096 \\ \hline
        Activation Function & GELU \\ \hline
        Layer Normalization Epsilon & 1e-05 \\ \hline
        Pooling Mode & Mean \\ \hline
        Problem Type & Single Label Classification \\ \hline
    \end{tabular}
    }
\end{table}
The phrase detection module efficiently classifies audio input into predefined categories, ensuring accurate identification of specified utterances, which is crucial for the text-dependent nature of the challenge. This dual-module approach, combining speaker verification and phrase detection, significantly enhances the overall effectiveness.


\section{Fine-Tuning, Training and Evaluation Strategy}

In this work, we focus on fine-tuning the NeXt-TDNN and ResNet TDNN models using a challenging dataset for speaker verification. The fine-tuning strategy emphasises a higher learning rate for the last layers of the models while maintaining a lower learning rate for the earlier layers, allowing for a more effective adaptation to the new data.

\subsection{Training Procedure}

The training process is structured as follows:

\begin{enumerate}
    \footnotesize 
    \item Initialise the model parameters using a random seed for training and load pre-trained weights for fine-tuning.
    \item Download and prepare the noise and room impulse response (RIR) datasets for data augmentation.
    \item Apply Voice Activity Detection (VAD) using the \texttt{pyannote/voice-activity-detection} module on every sample to remove silence at the beginning and end of each audio clip. This step is crucial, as silence does not provide useful information for speaker verification and could mislead the model by introducing artifacts from the same environment as the speakers.
    \item To enhance the robustness of our model, we employ various data augmentation techniques, summarized in Table~\ref{tab:data_augmentation}. These techniques introduce variability in the training data by simulating real-world conditions, such as noise and reverberation.

\begin{table*}[ht]
    \centering
    \caption{Summary of Data Augmentation Techniques}
    \begin{tabular}{|c|c|c|}
        \hline
        \textbf{Augmentation} & \textbf{Description} & \textbf{Parameters} \\
        \hline
        Noise Injection & Adds noise to clean signals & SNR: 0-15 dB \\
        \hline
        Reverberation & Applies room impulse responses & RIR annotations \\
        \hline
        Frequency Drop & Drops random frequency bands & 1-3 bands dropped \\
        \hline
        Time Drop & Removes random temporal chunks & Length: 1000-2000 ms, 1-5 drops \\
        \hline
    \end{tabular}
    \label{tab:data_augmentation}
\end{table*}

    \item Fine-tune both the NeXt-TDNN and ResNet TDNN models with the training dataset, leveraging the specified learning rates.
    \item Validate the model performance on a separate validation set after each epoch.
    \item Save checkpoints and logs periodically for further analysis and recovery.
\end{enumerate}
This training strategy is aimed at achieving better performance in speaker verification by effectively utilizing the challenge dataset while ensuring the models adapt properly to the new data.

\subsection{Evaluation Metrics}
To assess the performance of our models, we utilised the Minimum Detection Cost Function (MinDCF) and Equal Error Rate (EER) as our primary evaluation metrics. The MinDCF metric enables a comprehensive evaluation of the trade-off between false acceptance and false rejection rates, while the EER provides a straightforward comparison of the model's performance at varying threshold settings.
The main metric for the challenge is the normalised Minimum Detection Cost Function (DCF) as defined in SRE08. This detection cost function is defined as a weighted sum of miss and false alarm error probabilities.

\subsection{Evaluation Procedure}

The evaluation process is designed to ensure accurate and robust speaker verification, leveraging multiple enrollment files and a comprehensive score normalisation technique. The steps are as follows:

\begin{enumerate}
    \item \textbf{Enrollment Embedding Extraction}: For each speaker, the three enrollment utterances are concatenated following the application of VAD to remove any non-speech segments. This concatenation provides a more comprehensive representation of the speaker’s voice. Subsequently, a single embedding is extracted from the concatenated file, which serves as the enrollment speaker embedding.
    
    \item \textbf{Score Normalisation (S-norm)}: In the score normalisation step, embeddings from the training dataset are employed as cohort embeddings, without knowledge of the speaker identities in the evaluation set, ensuring unbiased normalisation. A cohort size of 10,000 embeddings is used to compute S-norm. This normalisation step refines the raw similarity scores by adjusting them according to the distribution of cohort embeddings, enhancing the robustness of the scoring mechanism.

    \item \textbf{Model Score Calculation and Fusion}: After extracting embeddings from both the enrollment and trial utterances, each Speaker Embedding Extractor, NeXt-TDNN, ResNet-TDNN, and EfficientNet, independently computes its similarity score. These scores are then converted into probability values in the range of 0 to 1. The final speaker verification score is obtained by applying a simple fusion of the normalised probabilities from the three models, allowing the system to combine their contributions in the final decision-making process.
    
    \item \textbf{Final Score Computation}: To compute the final decision score, the fused speaker verification score is multiplied by the probability output from the Phrase Classifier (PhC). This final score integrates both speaker identity verification and phrase content verification, ensuring that the system delivers a balanced decision in the context of text-dependent speaker verification tasks.

\end{enumerate}
This evaluation framework leverages the full potential of the enrollment data by using concatenated enrollment files, while the application of S-norm ensures that the scoring mechanism is resilient to variations in speaker embeddings. The fusion of multiple model outputs, combined with the incorporation of content verification via the PhC, significantly enhances the system’s reliability and generalisation across diverse speaker verification scenarios.

\subsection{System Configuration}

All models were fine-tuned and trained on a Linux system equipped with an NVIDIA 3090TI GPU. The training utilised only the challenge dataset, along with a subset of additional speakers: 128 speakers from the Mozilla Common Voice Farsi dataset and 210 speakers from the LibriSpeech dataset. These supplementary speakers were included to enhance the model's generalisation capabilities and to provide a richer diversity of speaker characteristics during training. The system was optimised for efficient utilisation of computational resources, ensuring that the large-scale training and evaluation tasks were completed within the constraints of the available hardware.

\section{Results}
\begin{figure*}[h!]
    \centering
    \includegraphics[width=0.7\textwidth]{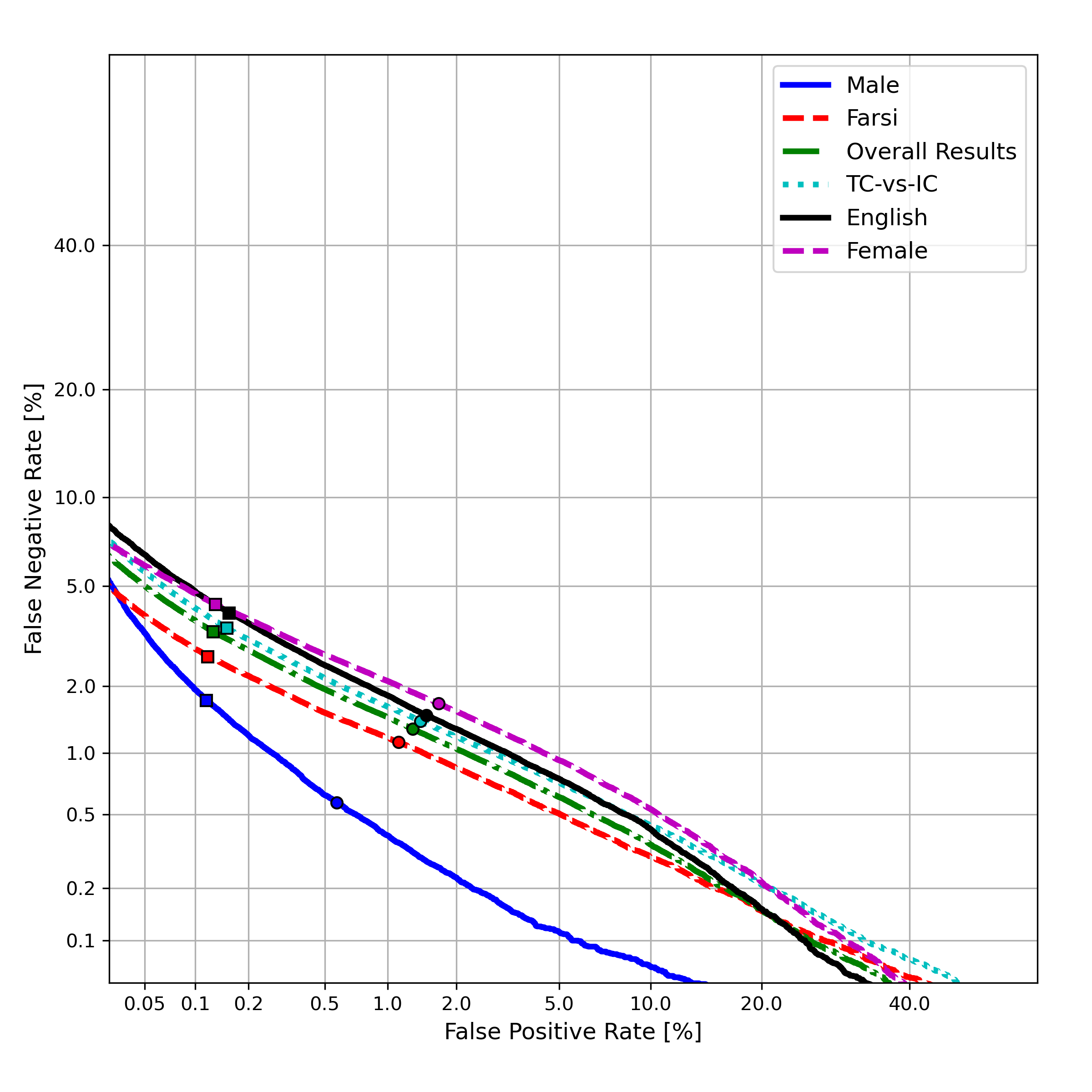}
    \caption{Detection Error Tradeoff (DET) curve comparing the performance across gender and language subsets.}
    \label{fig:det_curve}
\end{figure*}
The performance of the system in the 2024 Text-Dependent Speaker Verification (TdSV) Challenge is shown in Figure~\ref{fig:det_curve}, which presents the Detection Error Tradeoff (DET) curves for different demographic and linguistic groups. We achieved a minimum Detection Cost Function (MinDCF) of 0.0461 and an Equal Error Rate (EER) of 1.3\%.\\
As shown in the DET curves in Figure~\ref{fig:det_curve}, the performance across gender and language groups varies. The Male subset achieves the best performance, with a low False Negative Rate (FNR) across a wide range of False Positive Rates (FPR), as indicated by the blue solid line. In contrast, the Female subset, represented by the purple dashed line, shows a slightly higher FNR at similar FPR values, indicating a minor performance degradation for this demographic group.\\
The system also exhibits consistent performance across different language groups, with Farsi (red dashed line) and English (black solid line) showing similar trends, both slightly worse than the Male performance but well within acceptable ranges. Additionally, the performance of the system for text-constrained versus text-independent (TC-vs-IC) tasks, represented by the cyan dotted line, remains competitive with an overall steady performance. The overall results, indicated by the green solid line, demonstrate that the proposed system generalises well across the entire test set, maintaining robust verification capabilities across various demographic and linguistic conditions.

\section{Conclusion}
In this report, we presented a system for the 2024 TdSV Challenge, designed with a focus on robust and resource-efficient speaker verification through a multi-model ensemble. The system combines fine-tuned pretrained ResNet-TDNN and NeXt-TDNN models trained on the VoxCeleb dataset with a lightweight EfficientNet-A0 model trained on the challenge dataset to strengthen the ensemble approach.
Due to the limited time and resources available at the start of the challenge, this approach allowed us to make the best use of existing resources while still adapting effectively to the challenge data.\\
The proposed system combines three advanced neural network architectures for speaker embedding extraction and integrates a wav2vec 2.0 phrase verification module to address both speaker recognition and content verification tasks. By fusing the scores from the speaker and phrase verification modules, the system achieved strong generalisation and high accuracy across different demographic and linguistic groups, as shown by the DET curves.\\
Future work could involve refining model architectures to minimise the performance gap across gender and language subsets and exploring additional techniques for improving phrase verification accuracy, especially under variable challenge conditions.

\section{Acknowledgements}
We wish to express our sincere gratitude to the organisers of the TdSV Challenge 2024 for providing a valuable platform that facilitated the benchmarking and evaluation of our speaker verification system. We also acknowledge the developers of the DeepMine corpus for their contribution in providing the dataset crucial for this competition.


\bibliographystyle{ieeetr}
\bibliography{references}
\end{document}